# A Process To Support Cloud Release Preparation

**James J. Cusick, PMP**
IEEE Computer Society Member
Osaka, Japan
*j<dot>cusick<at>computer<dot>org*

**Abstract – This paper presents concepts and methods to support preparing software and system releases to production.**

## I. INTRODUCTION

With the intent of achieving better support systems and software quality in operations a method is described. This includes a checklist, a review process, and a dashboard for understanding status. Based on this approach release issues can be avoided and higher quality services can be delivered.

## II. ORR DESCRIBED

The essential form of the ORR (Operational Readiness Review) is shown in Figure 1. In the process a checklist (partly automated) is used during both development and service integration/deployment preparation to ensure and prepare technologies for usage. If they succeed in evaluation they require further mitigation and adjustment. Importantly, these checklists are always being improved as the technologies themselves are always changing and the team's knowledge about them is always improving. Further the possible failure scenarios may expand or change.

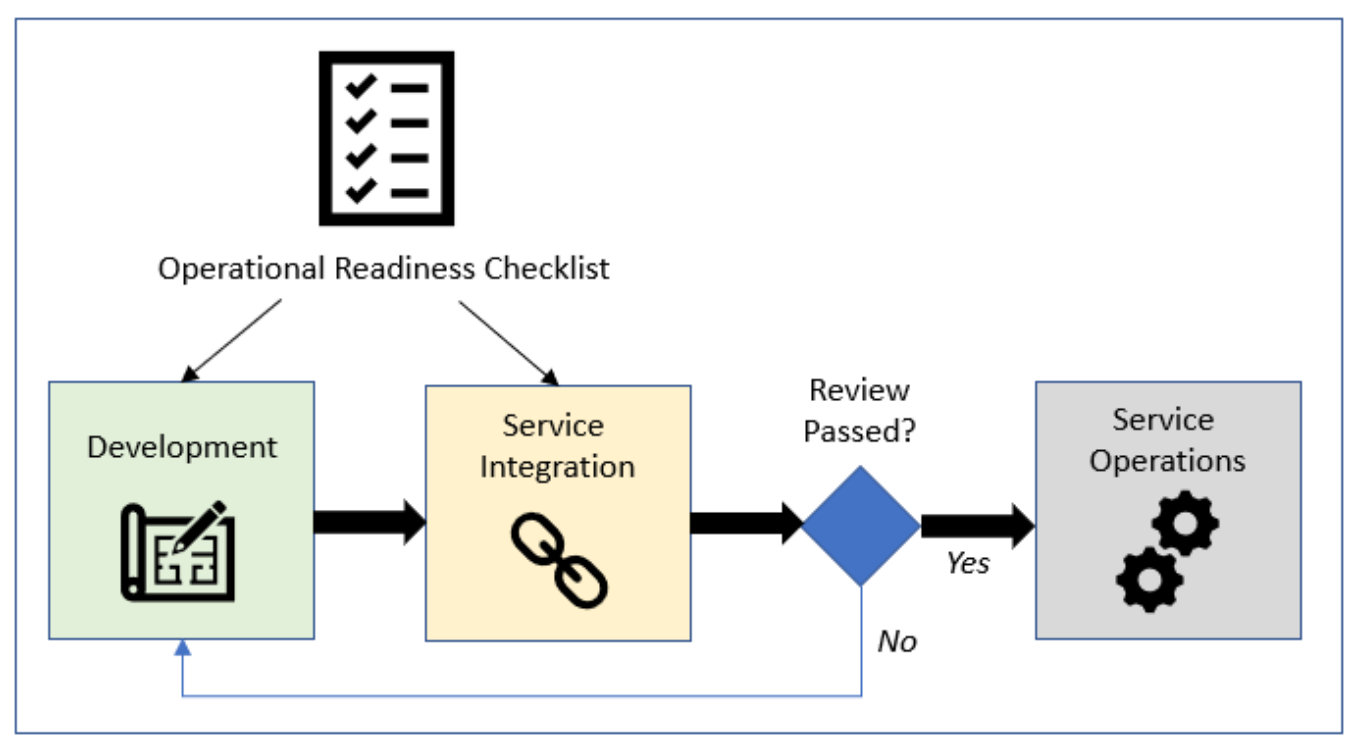

**Figure 1 - Essential ORR Process Framework [3]**

## III. ORR PROCESS BACKGROUND

Reviews like the ORR emerged from manufacturing and related disciplines with the intent to confirm the quality and preparedness of a given process. Similar practices were used in software and IT to review code and architectures. The ORR is also promoted by Project Management bodies like the PMI (Project Management Institute). The PMI takes the position that it is the project manager's responsibility to help guarantee a readiness program and perform required assessments. These actions are considered as necessary deliverables for any project [1].

## IV. CLOUD PROVIDERS AND ORR PROCESSES

With the increase in dominance of several major Cloud hosting providers in recent years most of them have also adopted a form of the ORR. For some companies they follow a standard ITIL model while others have customized their approach to suit their operational needs. In the case of Microsoft Azure's approach to the ORR their readiness process focuses on Security, Monitoring, High Availability, and Disaster Recovery [2].

For Amazon Web Services (AWS), a significant emphasis on the ORR to confirm applications are ready for production operations through their "Operational Excellence pillar" of their AWS Well-Architected Framework process is promoted [3]. The AWS ORR includes reviews of testing approach, monitoring, and an approach for measuring performance against the application's SLAs. Interestingly, AWS also ensures that applications can report data on service interruptions in a nod towards resiliency [4].

At Google, IT Operations are more commonly referred to as Site Reliability Engineering (SRE). Their pioneering work in largescale Cloud computing and operations has made them a trendsetter in this area. In the landmark book on SRE, Google lays out their computing environment management and engineering methods in detail [5]. The launch preparation method described in this book is like standard ORR methodologies. In addition, Google's approach plays a key role in their DevOps methods.

Finally, Hornsby [6] adds serval key considerations around the ORR for Cloud architects. He mentions that the ORR is rigorous and evidence based. It will also be specific to a company's environment, culture, and platforms. Regardless of this the ORR is always meant to uncover gaps and reduce risk to operations. This is why a broad set of contributors and stakeholders ought to be involved. Critically, Hornsby also makes the point that usage of an ORR is an exercise in incrementally improving the checklist as knowledge of the applications and computing assumptions evolve.

## V. Google, DevOps, and the ORR

One element of Google's launch checklist and PRR which distinguishes their approach from earlier ORR methods is the fact that Google is one of the foundational organizations for the emergence of DevOps. The SRE handbook documents a wide variety of now classical DevOps principles and methods which include [5]:

- Embracing Risk
- Automated Testing
- Capacity Management
- **PRR (Production Readiness Review)**
- Release Engineering (CI/CD)
- Operations Automation
- Monitoring
- Effective Incident Management

While many of these processes have been standard fare in the industry for decades, the bundling of them into a popularized movement has given them new attention. That includes the PRR, otherwise known as the ORR. Since Google's Site Reliability Engineering procedures are widely emulated the ORR has also gained attention in the DevOps community. This is crucial to the adoption of the ORR as today most IT environments strive to follow a DevOps mindset and methodology; however, they might be defined. The use of a PRR or ORR then, is not antithetical to DevOps but is clearly a core element for Google, AWS, and Azure. In the implementation of the Global ORR process described below the foundational inclusion of the PRR and ORR by the major Cloud operations providers and their DevOps processes was a key influence.

## VI. Developing a Global ORR Process

In order to improve system reliability and availability and to prevent unexpected environment anomalies and service disruption an effort was undertaken to develop and introduce a Global ORR process at a major information services company. This environment was truly global in that it operated in dozens of countries and serviced customers throughout the world. The scope included a portfolio of over 3,000 applications running on tens of thousands of Cloud based servers [7]. Heterogenous platforms and significant legacy and new development further characterized the environment. The Global IT process engineering team was tasked to develop the ORR approach at the request of the IT Operations Division CTO and CEO. IT Operations teams across multiple Divisions provided input to the ORR development and deployed the process as a new standard practice over time.

### *A. Process Goals*

As stated, the intention of the new Global ORR process was to enhance operational quality and system Availability to prevent service interruptions on behalf of customers. The specific goals informing the program included:

1. Drive uniform systems quality level with explicitly defined criteria.
2. Obtain Operational and Availability objectives and infuse continuous improvement throughout the systems operations process.
3. Provide transparency around support requirements and strengthen sharing of best practices across Global teams.

In addition to these primary drivers, it quickly became clear that the process would need to address the nature of the Agile and DevOps lifecycle models in use throughout the company. The implementation of the ORR process had to balance a comprehensive checklist approach with a DevOps approach which sees itself as highly Agile. This was one motivation for closely reviewing such Cloud based operational models as that from Google's SRE process which incorporates a PRR as standard practice while still retaining its DevOps approach.

### *B. ORR Process Development Steps*

After defining the need for a Global ORR process and setting in place the leadership support and objectives the development process was planned and initiated. The major steps of this development included the following:

1. **Initial Checklist Creation**: Within the company some Divisions did not have any pre-release checklist process. However, there were three internal groups with preexisting and custom operational checklist procedures in place. These were merged in order to seed the new Global ORR. This allowed for methods to be harvested and also began the buy-in process as those groups with operational checklists felt included in the development of the Global process.
2. **Industry Research**: As documented above, the team scanned the available literature for current practice in this area. Starting with the ITIL framework, carefully reviewing Google's SRE methods, and searching for other examples of ORR approaches all provided additional input to the definition of our new Global ORR approach and checklist.
3. **Executive Reviews**: As the process and checklist emerged into draft form it was reviewed with senior management. Their support for the process was unequivocal and they provided pointed feedback on improving the approach and their expectations on its usage.
4. **Pilot Phase**: The first deployment step was to select a small group of willing participants who had prior experience with PRRs or ORR-like preproduction checklist processes. This friendly and knowledgeable audience provided a testbed for the process as well as critical ideas on improving the process and the checklist itself. These learnings were used to retrofit the process artifacts and prepare for a broader rollout.

5. **Self-Reporting Phase**: Following the Pilot Phase all application teams in the initial population were required to participate in a “self-reporting” period. This allowed teams to begin picking up the process without tight oversight or data auditing.
6. **Full Deployment**: As the process was then stabilized and further improved, all application teams were required to participate through a defined governance process. Additionally, their readiness dashboards (see below) were reviewed by senior management as a condition of production release.
7. **Standard Global Practice**: Over the course of approximately 18 months the process went from concept to design to pilot to a required Global Practice. The checklist itself entered into planned and versioned releases to introduce improvements uniformly and globally without creating disruption in ORR usage. This steady state process is depicted in Figure 3. Also, governance migrated from the process development team to the Global IT PMO (Project Management Organization) group in coordination with the Architecture team and its review process.

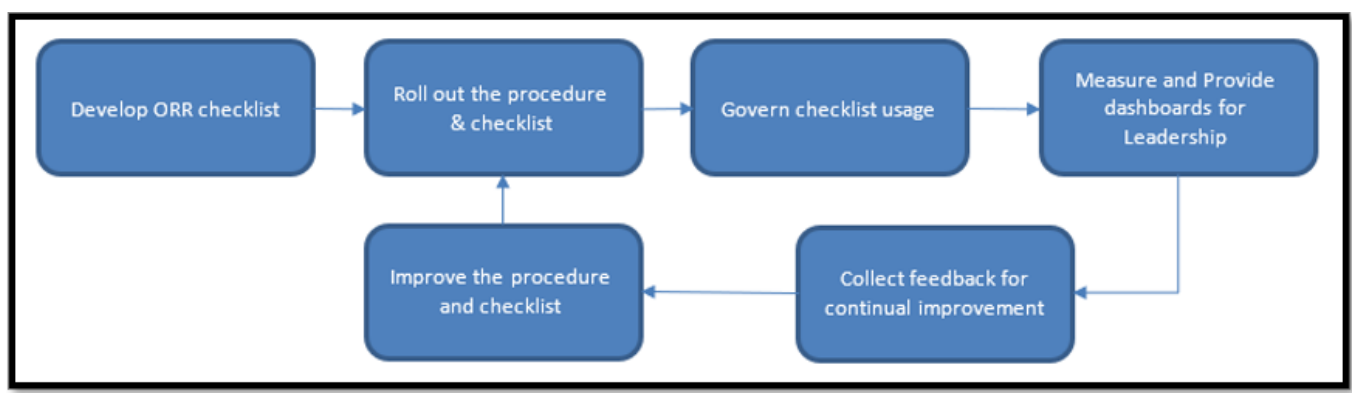

**Figure 2 - ORR Checklist Development and Governance Process Overview**

### *C. ORR Methodology and Tools*

In terms of the specific features or characteristics of this Global ORR process and checklist there are a few notable items listed below.

- The checklist itself was curated into a tuned, user verified, and proven list of just over 100 quality checkpoints. These categories included: Capacity, Performance, Touchpoints, Batch, Backup/Recovery, Support, Network/Firewalls, InfoSec, Cloud Computing, Monitoring, and DR.
- The process required a score of 100% against the applicable checkpoints. The questionnaire was largely automated and intelligent with a branching feature that allowed for differing checkpoint groups depending upon the architecture. The final score also required signoff from key stakeholders and senior management prior to production release.
- The ORR release preparation approach did not eliminate the Change Management process but strengthened it and helped reduce overall risk. Within the ORR process there was a defined Change Owner, supporting teams, Change Manager, and Leadership Authorizer.

**Sample ORR Executive Status Dashboard**

| ORR Checklist Categories | Regional Deployment Readiness | | | | |
|---|---|---|---|---|---|
| | Region 1 | Region 2 | Region 3 | Region 4 | Region 5 |
| Overall Score | 95% | 95% | 95% | 95% | 95% |
| Pre-conditions to ORR | 96% | 96% | 96% | 96% | 96% |
| Capacity Planning Readiness | 86% | 86% | 86% | 86% | 86% |
| Performances readiness | 100% | 100% | 100% | 100% | 100% |
| Batch Applications | N/A | N/A | N/A | N/A | N/A |
| Application Touchpoints | 80% | 100% | 100% | 100% | 100% |
| 3rd Party (Commercial Off-The-Shelf) | 100% | 100% | 100% | 100% | 100% |
| Backup / Recovery | 100% | 100% | 100% | 100% | 100% |
| Production Support | 100% | 88% | 88% | 88% | 88% |
| Networks & Firewalls | 75% | 75% | 75% | 75% | 75% |
| InfoSec & Malware | 86% | 86% | 86% | 86% | 86% |
| Storage | 100% | 100% | 100% | 100% | 100% |
| Servers & Hosts | 100% | 100% | 100% | 100% | 100% |
| Cloud Servers | N/A | N/A | N/A | N/A | N/A |
| Database | 100% | 100% | 100% | 100% | 100% |
| Data Streaming | N/A | N/A | N/A | N/A | N/A |
| Monitoring Tools | 100% | 100% | 100% | 100% | 100% |
| Disaster Recovery | 100% | 100% | 100% | 100% | 100% |
| Process | 100% | 100% | 100% | 100% | 100% |

**Figure 3 - Sample ORR Executive Dashboard. The design is inspired by the "Christmas Tree” operational safety readiness approach and visualization.**

- All systems, applications, and services of the organization were deemed within scope. The ORR became a mandated policy with senior management backing. Getting to full compliance took time and proceeded group by group and system type by system type but eventually reached universal adoption.
- Automation was applied at multiple points in the process to aid in adoption, data consistency, and the reduction of process overhead. Aside from the branching template to tailor responses to the application type or release, other key steps were also automated. These included the project team upload of the checklist to a central repository, automatic parsing of the checklists using PowerBI, and the automatic generation of an Executive dashboard.

## VII. Comparison with the Google Launch List

As part of the development and eventual communications around the new Global ORR process and checklist a comparison was conducted between the Google PRR Launch Checklist and the internal (or local) checklist under development. This assisted in confirming the comprehensiveness of the ORR implementation and its effectiveness as well as helping to demonstrate to internal teams that the Global ORR process compared well with state-of-the-art practices in major Internet technology companies who also employed DevOps.

As can be seen in Table 1 below, the two methods line up very closely. The only areas where Google is missing a category as compared with the Global ORR process are in Pre-requisites, Support, Database, and Process. However, we must keep in mind that the published Google checklist is abridged and is from 2005. This is to maintain some proprietary advantage.

Also, some of these categories are most likely covered within the SRE guide but are mentioned elsewhere. A good example is the Google PRR process itself which appears in the SRE book and calls for the use of the Launch Checklist. However, this is not mentioned explicitly on the Google Launch Checklist. Furthermore, nearly all Google applications use BigTable for data management, so this is an assumed implementation environment which has built in availability and scalability thereby perhaps reducing the need for a focused review section on the checklist.

**Table 1 - Google launch checklist items compared with the Global ORR checklist categories**

| ORR Category | Google | Global ORR |
|---|---|---|
| Pre-requisites | Partial | Y |
| Capacity | Y | Y |
| Performance | Y | Y |
| Batch | N | Y |
| Application | Y | Y |
| Third-Party | Y | Y |
| Backup | Y | Y |
| Support | Partial | Y |
| Network | Y | Y |
| Security | Y | Y |
| Storage | Y | Y |
| Hosting | Y | Y |
| Database | Indirectly | Y |
| Monitoring | Y | Y |
| DR | Y | Y |
| Process | N | Y |
| Failure Scenarios | Y | N |
| Automation | Y | N |

This puts the two checklists generally on par. However, of note are the two Google categories not covered directly by the Global ORR checklist. The first is the review of explicit failure scenarios and the handling of those scenarios by the system. This is a critical design step and readiness analysis which most operations teams might be advised to add. The other section directly called out be Google is automation. The ORR process described in this implementation touches on this tangentially but could be strengthened. An example of an automation step that would be helpful would cover firewall configuration. Instead of simply asking the question on whether the firewall ports have been assigned and opened, an executable script might be run in the actual computing environment to verify that the required ports are indeed open. Thus, as many of the ORR checkpoint questions as possible could be supported by both a review and a live capability confirmation. The results of the script could also feed into the dashboard automatically.

Additionally, there are some 70 topic areas to address in the abridged version of the Google checklist which may not line up exactly with those required by the Global ORR checklist under discussion. Some of these topics include Architecture, datacenter considerations, volumes, reliability, failover, and monitoring. Any of these checks might be considered for inclusion in a customized ORR. Readers can access the Google SRE guide for more information.

## VIII. Deployment Challenges

As with any change or transformation activity, developing the new process often turns out to be more straightforward than achieving successful adoption of that process. To do so often means teams and individuals must change their work habits, accept a new model of engineering or workflow, and typically requires a period of dialogue between those charged with introducing the change and those adopting the new procedures [8]. Achieving this type of process change across a large global organization is a project in itself and runs through phases of building the right climate for change, enabling the teams to succeed with the change, and finally implementing and sustaining the change [9]. In the case of the Global ORR process these steps were anticipated and significant effort applied to work through the ramp up of change and reach a status quo where the process became fully operational.

Along the way several key concerns were heard (some of them frequently) from the teams and organizations required to adopt and practice the Global ORR process.

- The first major complaint was that the ORR process (which was new to many project teams) added unnecessary overhead. While there was no doubt that the ORR represented a new quantum of work the response provided by the process team and the leadership team was that the upfront effort would pay off in less firefighting post-production. This was a difficult sell early on yet once teams began using the ORR they did see the benefits as they observed hidden preparation gaps and prevented major issues in the field.

## IX. Results and Discussion

Based on the development and use of this process and its supporting tools we have come to learn that the approach can yield benefits in both preventing release issues and in improving service quality. In many cases we have also seen improvements in delivery times once the teams learned to account for all the checkpoints in the matrix.

## X. Conclusions

We began this document by introducing how the ORR is defined within the industry. This included a review of how Cloud computing and DevOps driven providers incorporate ORRs. We also took a deep dive into the Google PRR (ORR) driven Launch Coordination Checklist. This included a discussion of the relation of such a checklist to DevOps. The details of the custom implementation of a new Global ORR for a specific organization were provided at length. Finally, we provided a side-by-side comparison of this locally customized ORR and Google readiness checklists and explored the challenges and results of developing and deploying an ORR process corporate wide.

In the end we can see that the intent, approach, and scope of the Google Launch Checklist and the Global ORR described here are very similar. There are some areas covered by one process which do not seem to be directly covered by the other and vice versa. This might suggest that implementation teams can continue to improve their methods by further review of industry practices.

Of note is that within a true DevOps environment like Google, a rigorous PRR or ORR method is expected. This should help IT Operations teams in finding adoption routes for the ORR in such process environments. Finally, as an industry standard method the ORR approach can be seen as a supporting model for achieving high levels of operational quality, reliability, and availability as in this case. Key success factors include Executive support and visibility to the ORR data, an active Change process to support adoption within the culture, and a universal application of the process to achieve even levels of higher systems quality and availability.

## XI. Acknowledgements

The development of this Global ORR process was dependent on the Executive management team of the organization profiled here as well as the process design team who developed and deployed the approach. Naturally, this work could not have been accomplished without the ultimate acceptance and cooperation of the hundreds of process users across the organization.

## XII. References

[1] Gardner, D. G., "*Operational readiness – is your system more "ready" than your environment?*" **Project Management Institute Annual Seminars & Symposium**, Nashville, TN. Newtown Square, PA: Project Management Institute, 2001.

[2] Azure Synapse, "*Operational Readiness Review*", **Synapse Success by Design: Implementation Success**, December 17, 2021.

[3] Eliot, Seth, et. al., "*Reliability Pillar: AWS Well-Architected Framework*", **Amazon Web Services**, July 2020.

[4] Cusick, James, "*Exploring System Resiliency and Supporting Design Methods*", **arXiv.org**, arXiv:0000 [cs.SE], September 2020.

[5] Murphy, Niall , et. al., **Site Reliability Engineering: How Google Runs Production Systems**, 1st Edition, O'Reilly, 2016.

[6] Hornsby, Adrian, "*Operational Readiness Review Template*", Towards Operational Excellence, **The Cloud Architect**, Nov 11, 2020.

[7] Cusick, James, "*Driving Enterprise Strategies with a Software Application Portfolio Management Initiative*", October 2020, **DOI: 10.13140/RG.2.2.12098.20167**.

[8] Cusick, James, "*Organizational Design and Change Management for IT Transformation: A Case Study*" **Journal of Computer Science and Information Technology**, 6(1), pp. 10-25, July 2018, DOI: 10.15640/jcsit.v6n1a2.

[9] Kotter, John P. & Cohen, Dan S., **The Heart of Change: Real-Life Stories of How People Change Their Organizations**, (1st edition). Harvard Business Review Press: Boston, MA, 2012.

## XIII. About the Author

James Cusick is a Principal IT Consultant & Researcher based in Japan advising select clients worldwide on IT Strategy, Cybersecurity, IT Operations, and Application Development. Previously James held leadership and technical roles in several companies including AT&T and Bell Laboratories. James is also currently a Board Trustee at the Henry George School of Social Science in New York researching Innovation, Technology, and Economics. Earlier James was an Adjunct Assistant Professor at Columbia University's Department of Computer Science. His publications include over 100 papers and two recent books on IT. James is a Member of the IEEE Computer Society and a certified PMP (Project Management Professional).